\documentclass[prb,aps,twocolumn,a4paper,showpacs,superscriptaddress]{revtex4-1}
\pdfoutput=1
\usepackage{graphicx}
\usepackage{amsmath}
\usepackage{amssymb}
\usepackage{bm}
\usepackage{dsfont}
\usepackage{graphicx}
\usepackage{subfigure}
\usepackage{lmodern}
\usepackage{url}
\usepackage[colorlinks=true,citecolor=blue,urlcolor=blue]{hyperref}

\newcommand{\re}{\mathrm{e}}
\newcommand{\ri}{\mathrm{i}}
\newcommand{\rA}{\mathrm{A}}
\newcommand{\rB}{\mathrm{B}}
\newcommand{\rJ}{\mathcal{J}}

\renewcommand{\ao}{\hat{a}}
\renewcommand{\aa}{\hat{a}^\dag}
\newcommand{\no}{\hat{n}}
\newcommand{\Ho}{\hat{H}}
\newcommand{\Uo}{\hat{U}}

\newcommand{\cfour}[4]{\aa_{#1} \aa_{#2} \ao_{#3} \ao_{#4} }

\newcommand{\hmn}[2]{h_{#1}({#2})}
\newcommand{\hmnconj}[2]{h_{#1}^*({#2})}

\newcommand{\la}{\langle}
\newcommand{\ra}{\rangle}

\newcommand{\ba}{{\bm a}}
\newcommand{\bk}{{\bm k}}

\newcommand{\br}{{\bm r}}
\newcommand{\bF}{{\bm F}}
\newcommand{\kkkk}{\left\{ \bk \right\}}

\newcommand{\be}{\begin{equation}}
\newcommand{\ee}{\end{equation}}
\newcommand{\bes}{\begin{eqnarray}}
\newcommand{\ees}{\end{eqnarray}}

\newcommand{\bose}{\mathrm{b}}
\newcommand{\fermi}{\mathrm{f}}
\newcommand{\kin}{\mathrm{kin}}
\newcommand{\inter}{\mathrm{int}}

\usepackage[x11names,dvipsnames]{xcolor}
\definecolor{vured}{RGB}{137,28,46}
\definecolor{vugrey}{RGB}{197,197,197}
\definecolor{vugreen}{RGB}{34,139,34}

\begin{document}

\title{The role of real-space micromotion for bosonic and fermionic \\ Floquet fractional Chern insulators}
\date{14 April 2015}

\author{Egidijus Anisimovas}
\email{egidijus.anisimovas@ff.vu.lt}
\affiliation{Institute of Theoretical Physics and Astronomy, Vilnius University,
A. Go\v{s}tauto 12, LT-01108 Vilnius, Lithuania}
\affiliation{Department of Theoretical Physics, Vilnius University, 
Saul\.{e}tekio 9, LT-10222 Vilnius, Lithuania}
\author{Giedrius \v{Z}labys}
\affiliation{Institute of Theoretical Physics and Astronomy, Vilnius University,
A. Go\v{s}tauto 12, LT-01108 Vilnius, Lithuania}
\affiliation{Department of Theoretical Physics, Vilnius University, 
Saul\.{e}tekio 9, LT-10222 Vilnius, Lithuania}
\author{Brandon M. Anderson}
\affiliation{James Franck Institute, Enrico Fermi Institute and Department of Physics, 
University of Chicago, Chicago, IL 60637, USA}
\affiliation{Joint Quantum Institute, University of Maryland, College Park, MD 20742, USA}
\author{Gediminas Juzeli\={u}nas}
\affiliation{Institute of Theoretical Physics and Astronomy, Vilnius University,
A. Go\v{s}tauto 12, LT-01108 Vilnius, Lithuania}
\author{Andr\'e Eckardt}
\email{eckardt@pks.mpg.de}
\affiliation{Max-Planck-Institut f\"ur Physik komplexer Systeme, 
\mbox{N\"othnitzer Stra\ss e 38, 01187 Dresden, Germany}}

\begin{abstract}
Fractional Chern insulators are the proposed phases of matter mimicking the physics of fractional 
quantum Hall states on a lattice without an overall magnetic field. The notion of Floquet 
fractional Chern insulators refers to the potential possibilities to generate the underlying 
topological bandstructure by means of Floquet engineering. In these schemes, a highly controllable 
and strongly interacting system is periodically driven by an external force at a frequency such 
that double tunneling events during one forcing period become important and contribute to shaping 
the required effective energy bands. We show that in the described circumstances it is necessary 
to take into account also third order processes combining two tunneling events with interactions. 
Referring to the obtained contributions as micromotion-induced interactions, we find that those 
interactions tend to have a negative impact on the stability of of fractional Chern insulating 
phases and discuss implications for future experiments.
\end{abstract}


\maketitle

\section{Introduction}

Chern insulators\cite{haldane88} form a primary and most widely studied class of more general 
topological insulators,\cite{hasan10,qi11} proposed in condensed matter settings, in particular, 
heterostructures\cite{bernevig06,hsieh08,konig12} and graphene.\cite{kane05a,kane05b} 
They are characterized by topological Bloch bands, that is, energy bands that give 
rise to a quantized Hall conductivity, when filled completely in a band-insulating state. The 
contribution of each band to the Hall conductivity is identified by an integer topological 
index, known as the Chern number. From this point of view topological bands can be thought of 
as generalizations of the Landau level.\cite{klitzing86} However, they can be realized in a 
broad variety of physical settings independent of the requirement to have electrically charged 
particles coupled to an intense uniform magnetic field. The analogy between a Landau level and 
a Chern band hints at the idea of \emph{fractional} Chern insulators 
(FCI).\cite{sheng11,neupert11,regnault11,wang11,wu12,bergholtz13,parameswaran13,grushin14}
In this case, in addition to an isolated energy band characterized by a nonvanishing integral 
of the Berry curvature (which defines the Chern number) one also needs strong particle 
interactions to form the collective states analogous to (and presumably richer than) the usual 
fractional quantum Hall states.

It turned out that a powerful method to produce the desired topological bandstructures is
\emph{Floquet engineering}. This form of quantum engineering is based on the fact that the 
dynamics of a time-periodically driven quantum system, a so-called Floquet system, is (apart 
from a periodic micromotion) captured by a time-independent effective Hamiltonian. Properties 
of the effective Hamiltonian can be engineered by tailoring a 
suitable driving protocol. Floquet engineering has been very successfully applied to quantum 
systems of ultracold atoms in periodically driven optical 
lattices.\cite{eckardt05,lignier07,sias08,zenesini09,alberti09,eckardt10,haller10,struck11,kolovsky11,aidelsburger11,hauke12,
struck12,struck13,aidelsburger13,miyake13,jotzu14,goldmanreview14}
These systems are particularly suitable for such control schemes due to their high degree of 
isolation from the environment and versatile possibilities for controlling parameters in a 
time-dependent fashion during the experiment. In recent years, several schemes have been proposed 
where periodic driving is employed for the realization of topologically nontrivial effective 
bandstructures in lattice systems, which in the absence of the driving are topologically 
trivial. These schemes can be divided into two classes: (i) methods working at high driving 
frequencies\cite{sorensen05,bermudez11,kolovsky11,hauke12,aidelsburger13,miyake13,aidelsburger14,baur14} 
and relying on averaging the driven Hamiltonian over a period, and (ii) methods working at 
intermediate driving frequencies\cite{oka09,kitagawa10,perez14,rechtsman13,jotzu14} and going
beyond the time-averaged description. For the latter, the term 
\emph{Floquet topological insulator} has been coined.\cite{lindner11,dora12} 

The high-frequency schemes have been proposed and implemented in the context of ultracold atomic 
quantum gases in optical 
lattices\cite{sorensen05,kolovsky11,aidelsburger11,struck12,hauke12,aidelsburger13,miyake13,struck13,aidelsburger14} 
and trapped ions.\cite{bermudez11} Here, 
static and time-periodic potentials are combined in such a way that the wavefunction acquires 
time-periodic relative phases on neighboring lattice sites. When averaged over one driving 
period, these phases resemble the nontrivial phases induced by a magnetic field. 
These schemes work at large frequencies, since it is assumed that the driving period $T$ 
determining the phase modulation is short compared to the tunneling time. The topologically 
nontrivial effective bandstructure of such a system has been recently probed 
in a square optical lattice.\cite{aidelsburger14}

The Floquet topological insulator schemes \cite{oka09,kitagawa10,lindner11} were originally 
proposed in the context of condensed matter systems, considering irradiated graphene or semiconductor 
heterostructures, and are based on a different principle. Here, the effective Hamiltonian 
acquires new terms that describe tunneling between next-nearest neighbors and open a topological 
gap in the bandstructure. These new terms are related to second-order processes, where a particle 
tunnels twice during one driving period. Obviously they are significant only at intermediate 
driving frequencies, with the driving period being comparable to (or at most moderately shorter 
than) the tunneling time. This is the regime where particles undergo a significant periodic 
real-space micromotion in response to the driving. Signatures of such Floquet topological 
bandstructures have been observed in optical waveguides\cite{rechtsman13} and with ultracold 
fermionic atoms in a shaken honeycomb lattice.\cite{jotzu14}

As a natural next step, recently it has been proposed to stabilize a fractional topological 
insulator phase in such a Floquet topological bandstructure, referred to as the Floquet 
fractional Chern 
insulator.\cite{grushin14} The scheme is based on fermions in a circularly driven honeycomb 
lattice, which acquires a topologically nontrivial effective bandstructure.\cite{oka09} For 
their analysis, the authors employed a high-frequency approximation\cite{goldman14,itin14,eckardt15} 
of the effective Hamiltonian, including terms up to the second order. On this level of approximation, 
the nontrivial tunneling terms opening the topological single-particle bandgap are included, 
but no corrections to the interaction terms appear.\cite{eckardt15, grushin15} The exact 
diagonalization of the approximate effective Hamiltonian for small systems suggested that a 
topologically ordered state can be stabilized at a band filling of $1/3$.

The realization of such a topologically ordered many-body Floquet state is, however, challenged 
in various ways. One difficulty concerns the preparation of the state both in open condensed 
matter systems as well as in isolated cold-atom systems. Open Floquet systems assume steady 
states, which are generally quite different from 
equilibrium states\cite{bluemel91,kohler97,breuer00,hone09,ketzmerick10,vorberg13,seetharam15,iadecola15} 
and in an isolated system the desired state has to be reached by an adiabatic passage starting 
from the undriven ground states. A second concern is that excitations to higher-lying bands, 
which are not included in the tight-binding description, via multi-``photon'' transitions may become 
relevant on the time scale of the experiment. Finally, also the impact of interactions beyond 
the second-order high-frequency approximation can be a relevant issue. Such ``residual'' 
interactions enter on two different levels, they cause heating and they lead to higher-order 
corrections to the approximate effective Hamiltonian. Interaction-induced heating corresponds 
to processes that can be viewed as the resonant creation of collective excitations of the 
effective Hamiltonian in high-frequency approximation. Such processes are not captured within 
the high-frequency expansion, they indicate that such an expansion cannot be expected to converge 
for an interacting system\cite{eckardt15} and they are expected to eventually drive the system 
towards an infinite-temperature regime.\cite{lazarides14,dalessio14} 
A perturbative argument suggests that the rate of such detrimental heating will decrease 
exponentially with increasing driving frequency (as the order in which the corresponding processes 
appear increases with the driving frequency).\cite{eckardt08}
Apart from heating, interactions will also lead to corrections appearing in higher orders of 
the high-frequency expansion, with the leading correction appearing in third order.\cite{eckardt15}

In this paper we investigate the impact of leading interaction correction to the approximate 
effective Hamiltonian. We do not address the issues of preparation and heating due to either 
multi-photon interband transitions or the resonant excitation of collective excitations. 
While it is rather clear that heating is a detrimental effect, it is 
an interesting question without an \emph{a priori} obvious answer, whether the interaction corrections 
will tend to stabilize or to destabilize a Floquet fractional Chern insulator state. Even 
though the leading interaction corrections appear in third order only, they are still relevant 
for the required intermediate driving frequencies. First of all, if interactions are strong 
compared to tunneling, a third-order interaction correction can be comparable to a 
second-order kinetic term. And second, because the correction has to be compared with the 
tiny manybody gap that protects the ground-state manifold of the effective Hamiltonian from 
excited states. The origin of the interaction corrections to be investigated here is a 
significant real-space micromotion at intermediate driving frequencies. A particle at a certain 
lattice site will explore also neighboring sites during one driving period. This real-space 
micromotion generates new effective interactions at distances longer than those of the bare 
interactions characterizing the undriven model.

For our study, we use exact diagonalization for small systems, taking into account the leading 
interaction correction. In addition to a model of spin-polarized fermions with nearest-neighbor 
interactions (which includes the ground states found in Ref.~\onlinecite{grushin14}) we also 
consider a system of spinless bosons with onsite interactions. The latter is particularly 
interesting for experiments with ultracold atoms in shaken optical lattices, and is 
typically governed by onsite interactions. In both models we find that taking into account 
interaction corrections tends to destabilize topologically ordered fractional Chern states.
Thus the realization of Floquet fractional Chern insulator states seems rather challenging
and it might be more promising to consider high-frequency schemes or schemes involving internal 
degrees of freedom.\cite{goldmanreview14,dalibard11}

The bulk of the presented material is split between two large sections, each subdivided into 
three subsections. Sec.~\ref{sec:mod} discusses the model, starting with the description of the
lattice and the shaking protocol, and later proceeding to the Floquet analysis, the underlying 
single-particle problem and the role of interactions. Section~\ref{sec:res} focuses on results,
and encompasses 
description of the numerical procedure, the obtained manybody bandstructures and quasihole 
spectra, characterizing the fractional Chern insulating states. Finally, we conclude with a 
brief summarizing Sec.~\ref{sec:con}.

\section{Model}\label{sec:mod}

\begin{figure}[ht]
\begin{center}
\includegraphics[width=50mm]{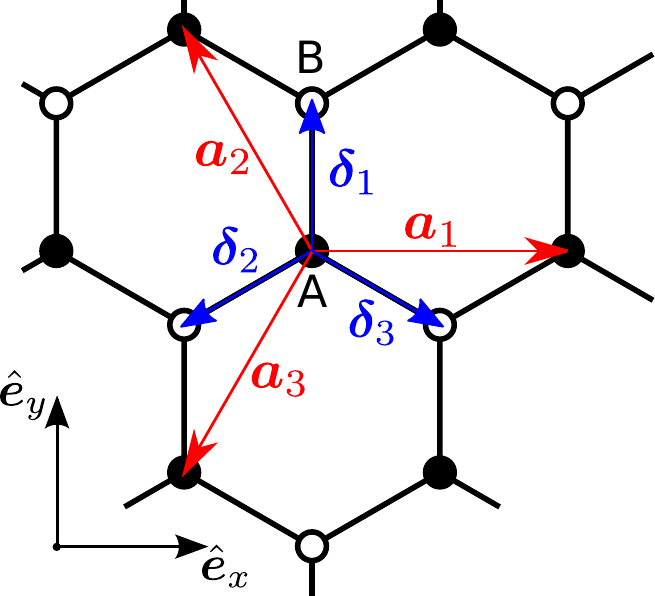}
\caption{\label{fig:latt}Honeycomb lattice as a triangular Bravais lattice with a two-site basis. 
The inequivalent sites A and B are marked with filled and open dots, respectively. Nearest 
neighbors are connected by the vectors $\bm\delta_{\mu}$, while the set of next-nearest 
neighbor vectors $\ba_j$ also defines the elementary translations.}
\end{center}
\end{figure}

The model proposed by Grushin et al.~\cite{grushin14} is based on a honeycomb lattice sketched in 
Fig.~\ref{fig:latt}, and consisting of two intertwined triangular sublattices A and B. For the 
sake of further reference, Fig.~\ref{fig:latt} defines the vectors 
\begin{equation}
  \bm{\delta}_1 = a \, \hat{\bm e}_y, \qquad
  \bm{\delta}_{2|3} = \mp \tfrac{\sqrt{3}}{2} a \, \hat{\bm e}_x 
  + \tfrac{1}{2} a \, \hat{\bm e}_y,
\end{equation}
connecting a given site A to its three nearest neighbors (NN) on sublattice B, with $a$ denoting 
the nearest-neighbor distance, and $\hat{\bm e}_{x,y}$ being the unit coordinate vectors. Likewise,
the vectors $\pm \bm{a}$ with
\begin{equation}
  \bm{a}_1 = \sqrt{3} a \,\hat{\bm e}_x, \qquad
  \bm{a}_{2|3} = - \tfrac{\sqrt{3}}{2} a \, \hat{\bm e}_x 
  \pm \tfrac{3}{2} a \, \hat{\bm e}_y,
\end{equation}
connect a given site of either type to its six next-nearest neighbors (NNN) belonging to the same 
sublattice. Note that NNN hopping transitions naturally separate into two classes. Transitions 
starting from a site A (resp.\ B) in the direction of $\bm{a}_j$ correspond to counterclockwise 
(resp.\ clockwise) motion around the hexagonal cell, and the opposite statement applies to 
transitions along $-\ba_j$. The vectors $\ba_1$ and $\ba_2$ are also taken to define the Bravais 
lattice $\br = n_1 \ba_1 + n_2 \ba_2$, while $\ba_3 = - \ba_1 - \ba_2$ is linearly dependent.

The honeycomb lattice is subjected to a circular time-periodic force of fixed magnitude $F$,
whose direction rotates in the ($x$-$y$) lattice plane at a constant frequency $\omega$, thus
\begin{equation}
  \bF (t) = F \sin\omega t \, \hat{\bm{e}}_x - F \cos\omega t \, \hat{\bm{e}}_y.
\end{equation}
This force results from either the circular shaking\cite{eckardt10,jotzu14,zheng14}
of an optical lattice or, in the case of charged 
particles, from irradiation by circularly polarized light.\cite{oka09} A suitable gauge 
transformation~\cite{eckardt15,goldmanreview14} (corresponding to transition to the comoving
frame of reference) restores the translational invariance of the Hamiltonian, 
which is then represented in the tight-binding form as a sum of the time-periodic kinetic 
and time-independent interaction parts, 
\begin{equation}
  \Ho(t)=\Ho_{\kin}(t)+\Ho_{\inter}.
\end{equation}

The kinetic Hamiltonian reads
\begin{equation}
  \label{eq:Hkin}
  \Ho_{\kin}(t) =-\sum_{i\in \rA} \sum_{\mu=1}^3 J_\mu(t) \, \ao_{i+\mu}^\dag \ao_{i} + h.c.,
\end{equation}
here, the operator $\aa_{i}$ creates a spinless particle (fermion or boson) on lattice site 
$i$. The index $\mu$ labels the three distinct directions connecting any given site of 
sublattice A to its nearest neighbors belonging to sublattice B, and denoted by the shorthand 
label ``$i+\mu$'' with $i \in \rA$. Extension to spinful particles, although not necessary for 
the purposes of the current presentation, is straightforward.
Equation~(\ref{eq:Hkin}) describes the nearest-neighbor tunneling kinetics. The tunneling matrix 
elements along the three different directions acquire their time dependence due to the circular 
driving, and thus are time-periodic with uniformly distributed relative phases, i.e., 
\begin{equation}
  J_\mu(t)=J(t - \varphi_{\mu} / \omega), \quad\text{with}\quad 
  \varphi_{\mu} = \frac{2\pi}{3}(\mu - 1). 
\end{equation}
Having in mind 
applications to ultracold atoms in a circularly shaken optical lattice or graphene electrons 
irradiated by circularly polarized light, we write 
\begin{equation}
  J(t) = J\exp(i\alpha\sin\omega t),
\end{equation}
with 
\begin{equation}
  \label{eq:alpha}
  \alpha = \frac{F a}{ \hbar\omega}
\end{equation}
denoting the dimensionless shaking strength. 

In bosonic quantum gases contact interactions may be assumed, therefore, the interaction 
Hamiltonian is written in the standard Bose-Hubbard form
\begin{equation}
  \label{eq:Hintb}
  \Ho_{\inter}^{(\bose)} 
  = \frac{U}{2}\sum_{i \in \rA, \rB} \no_{i} \left(\no_i -1 \right), 
\end{equation}
with the operator $\no_{i}=\aa_{i}\ao_{i}$ measuring the particle number on a given site.
In contrast, for spinless fermions the onsite interaction term vanishes due to the Pauli exclusion 
principle, and repulsion between pairs of particles occupying neighboring sites must be taken 
into account, thus
\begin{equation}
  \label{eq:Hintf}
  \Ho_{\inter}^{(\fermi)} = V \sum_{i\in \rA } \sum_{\mu=1}^3 \no_{i}\no_{i+\mu}.
\end{equation}
Therefore, depending on the particle statistics, we model the interactions using one of the
two alternative forms given by either Eq.~(\ref{eq:Hintb}) or (\ref{eq:Hintf}). Note also, 
that in both cases the interaction Hamiltonian depends only on densities, and thus remains
static also in the presence of periodic driving. Our choice of treating spinless, that is 
spin-polarized, fermions is motivated not only by simplicity, but also by the observation 
of Ref.~\onlinecite{grushin14} that the relevant Floquet fractional Chern insulating states 
are ferromagnetic. 

\subsection{Floquet analysis}

As discussed in the Introduction, Floquet Chern insulators by construction require intermediate 
driving frequencies on the order of the tunneling strength or just moderately larger. This 
observation identifies the dimensionless inverse shaking frequency
\begin{equation}
  \beta = \frac{J}{\hbar\omega}
\end{equation}
as the series-expansion parameter that classifies the successive contributions to the effective 
Hamiltonian, and in turn, to the ensuing physics. In further study we focus on the interval
$0.1 \leqslant \beta \leqslant 0.5$. At the lower limit of this range, one crosses over to the 
high-frequency regime where the effective Hamiltonian is adequately represented by the time 
average of the driven Hamiltonian while the relevance of higher-order contributions fades away. 
In the opposite limit, for values of $\beta$ exceeding one-half the representation of the 
effective Hamiltonian in terms of a $\beta$-series expansion in no longer reliable.

A powerful description of periodically driven quantum systems\cite{eckardt15,goldman14,rahav03,gdac15,itin14} 
relies on the factorization of the quantum-mechanical evolution operator according to
\begin{equation}
  \Uo (t_2, t_1) = \Uo_F (t_2) \, \re^{-\ri \Ho_F (t_2 - t_1) / \hbar} \, \Uo_F^{\dag} (t_1),
\end{equation}
thus separating micromotion described by the time-periodic unitary micromotion operator 
$\Uo_F (t)$ from the long-term dynamics captured by the effective Hamiltonian $\Ho_F$.
A consistent high-frequency approximation to the effective Hamiltonian $\Ho_F$ is constructed 
in Refs.~\onlinecite{eckardt15,goldman14,rahav03} following different approaches
and resulting in 
equivalent\footnote{Formula (C.10) in Ref.~\onlinecite{goldman14} differs from our Eq.~(\ref{eq:hfc})
by an apparently spurious factor of 2 in front of the last triple commutator which, however, 
does not contribute in the present case.} series representation in powers of $\omega^{-1}$
\begin{equation}
  \Ho_F = \Ho_F^{(1)} + \Ho_F^{(2)} + \Ho_F^{(3)} + \cdots,
\end{equation}
where
\begin{widetext}
\begin{subequations} \label{eq:hf}
\begin{align}
  \Ho_F^{(1)} &= \Ho_{0}, \label{eq:hfa}\\
  \Ho_F^{(2)} &= \sum_{m=1}^{\infty} \frac{1}{m \hbar \omega} 
    \left[\Ho_m, \Ho_m^{\dag} \right], \\
  \label{eq:hfc}  
  \Ho_F^{(3)} &= \frac{1}{2 (\hbar\omega)^2} \sum_{m=1}^{\infty}
    \frac{ [\Ho_m, [\Ho_{0}, \Ho_m^{\dag}]]}{m^2} 
    + \frac{1}{3(\hbar\omega)^2} \sum_{\substack{m,m'=1 \\ m' \neq m}}^{\infty} 
    \frac{[\Ho_{-m'},[\Ho_{m'-m}, \Ho_{m}]] - [\Ho_{m'},[\Ho_{-m'-m}, \Ho_{m}]]}{mm'} + h.~c.,
\end{align}
\end{subequations}
\end{widetext}
are expressed in terms of Fourier components of the driven Hamiltonian 
\begin{equation}
  \Ho (t) = \sum_{m=-\infty}^{\infty} \Ho_m \, \re^{\ri m\omega t}.
\end{equation}
It is useful to note the hermiticity condition $\Ho_{-m} = \Ho_m^{\dag}$.

\begin{figure}[ht]
\begin{center}
\includegraphics[width=84mm]{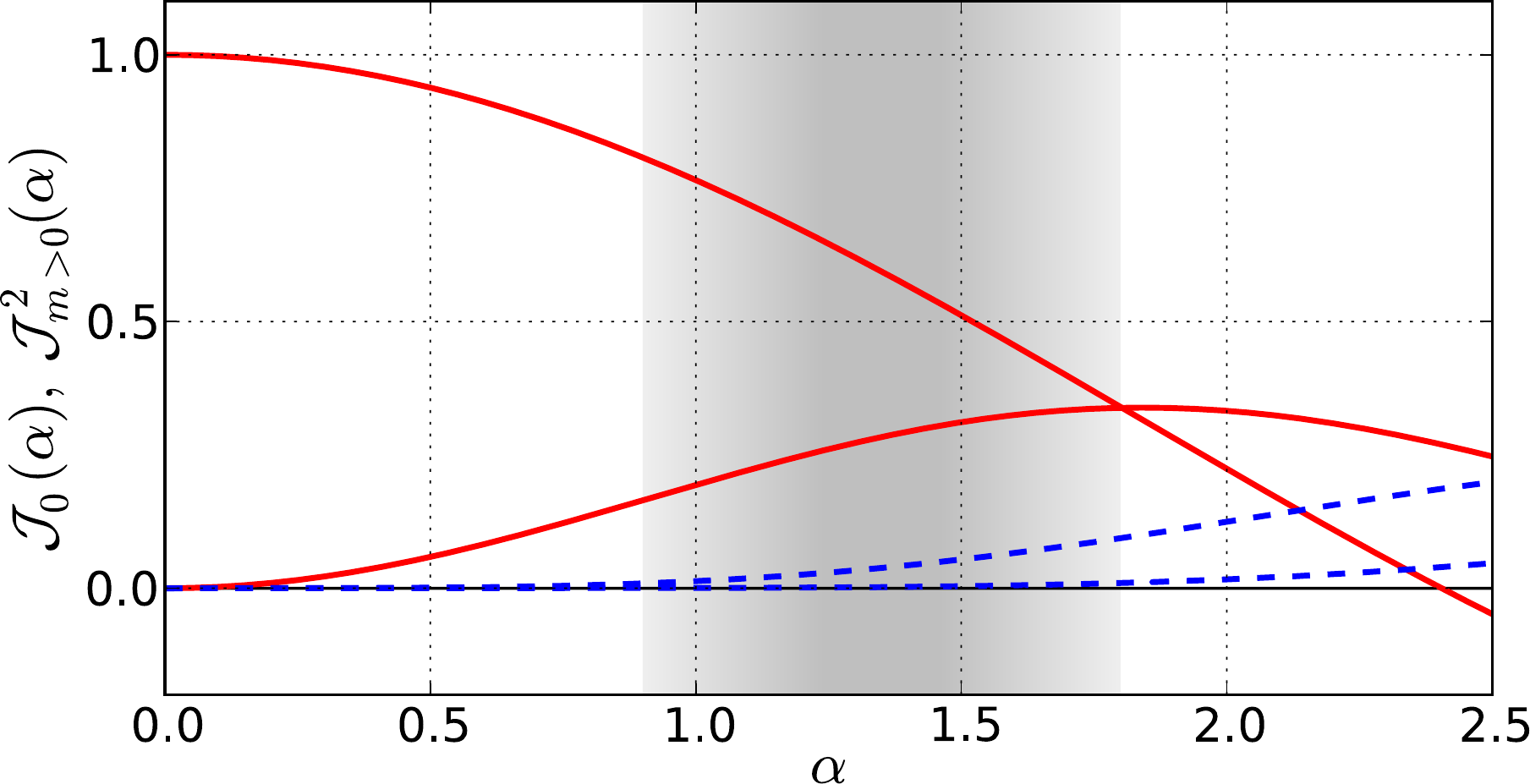}
\caption{\label{fig:bess}Comparison of the importance of various Fourier harmonics. The 
shaded background shows the range of the dimensionless shaking amplitudes $\alpha$ considered 
in the calculations. The red lines depict the behavior of $\mathcal{J}_0 (\alpha)$ and 
$\mathcal{J}_1^2 (\alpha)$, setting the scale of, respectively, the static components of the 
Hamiltonian and the first harmonic. The dashed blue lines show the behavior of the (omitted)
subleading contributions $\mathcal{J}_{2,3}^2 (\alpha)$.}
\end{center}
\end{figure}

For the purposes of the current application, the general expressions~(\ref{eq:hf}) can be 
simplified in two aspects. First, we note that the time dependence of the driven Hamiltonian 
stems entirely from the time-dependent hopping amplitudes described by the Fourier expansion 
\begin{equation}
  J_{\mu} (t) = \sum_{m=-\infty}^{\infty} J \rJ_m(\alpha) \,
  \re^{-\ri m\varphi_{\mu}} \, \re^{\ri m\omega t},
\end{equation}
with Bessel functions of the first kind (and order $m$), 
denoted here by $\rJ_m$. Anticipating moderate shaking strengths $\alpha$ [see Table~\ref{tab}]
we note that the Fourier series of the Hamiltonian converge very fast since contributions 
originating from the $m$th harmonic with $m > 0$ include a prefactor 
$\rJ_m^2(\alpha) \sim \alpha^{2|m|}$ (see Fig.~\ref{fig:bess}). For this reason, we truncate the 
Fourier expansion and include only the terms with $|m| \leqslant 1$. Consequently, non-zero
Fourier components of the driven Hamiltonian are
\begin{subequations}
\begin{align}
  \Ho_0 &= \Ho_{\inter} - \sum_{i \in \rA} \sum_{\mu} J \rJ_0 (\alpha) 
    \left[\aa_{i+\mu} \ao_{i} + \aa_{i} \ao_{i+\mu} \right], \label{eq:hamfd}\\
  \Ho_{\pm 1} &= \mp \sum_{i \in \rA} \sum_{\mu} J \rJ_1 (\alpha) \, \re^{\mp\ri\varphi_{\mu}} 
    \left[\aa_{i+\mu} \ao_{i} - \aa_{i} \ao_{i+\mu} \right],
\end{align}
\end{subequations}
where the interaction Hamiltonian $\Ho_{\inter}$ is given by either Eq.~(\ref{eq:Hintb}) or 
Eq.~(\ref{eq:Hintf}) for bosonic and fermionic systems, respectively. 

Secondly, focusing on the interplay of micromotion and interactions we take into account the
leading third-order interaction correction while neglecting purely kinetic terms of third order.
Thus, the expansion of the effective Hamiltonian considered in our work reads
\begin{subequations}
\label{eq:hamf}
\begin{align}
  \Ho_F^{(1)} &= \Ho_{0}, \label{eq:hamfa} \\
  \Ho_F^{(2)} &= \frac{1}{\hbar\omega} \left[\Ho_1, \Ho_1^{\dag} \right], \label{eq:hamfb}\\
  \Ho_F^{(3)} &= \frac{1}{2 (\hbar\omega)^2} 
    \left[ \Ho_1, \left[ \Ho_{\inter}, \Ho_1^{\dag} \right] \right] + {h. c.}. \label{eq:hf3fin}
\end{align}
\end{subequations}

\subsection{Single-particle spectrum}

The first-order contribution to the effective Hamiltonian given by Eq.~(\ref{eq:hamfa}) is
obtained by averaging the driven Hamiltonian $\Ho(t)$ over a period. This leads to the result 
given by Eq.~(\ref{eq:hamfd}) and featuring the well known 
renormalization of NN hopping amplitude according to the prescription $J \to J \rJ_0 (\alpha)$,
which has been extensively exploited to demonstrate a number of interesting physical effects 
in cold atom settings, including the superfluid-Mott insulator phase 
transition.\cite{eckardt05,zenesini09}

The second-order Floquet contribution is expressed by the commutator in Eq.~(\ref{eq:hamfb}),
and physically corresponds to two subsequent tunneling processes of a particle during one 
driving period. The basic commutation relation 
$[\aa_k \ao_{\ell}, \aa_m \ao_n] = \delta_{\ell m} \aa_k \ao_n - \delta_{kn} \aa_m \ao_{\ell}$
is valid for particles of either statistics and combines two successive NN tunneling events 
into an effective NNN transition. Thus,
\begin{equation}
\label{eq:hf2}
  \Ho_F^{(2)} = - \sum_{\la\!\la ij \ra\!\ra} 
  J_{\la\!\la ij \ra\!\ra}^{(2)} \aa_i \ao_j,
\end{equation}
with the sum running over all next-nearest neighbor connections, and
\begin{equation}
  J_{\la\!\la ij \ra\!\ra}^{(2)} = \pm \ri \sqrt{3} \, \beta J \rJ_1^2 (\alpha).
\end{equation}
The plus (minus) sign applies to clockwise (counterclockwise) transitions around the hexagonal 
unit cell. Note that NNN transitions in Eq.~(\ref{eq:hf2}) are characterized by purely imaginary 
amplitudes.

The role of the second-order contribution to the effective Hamiltonian in Floquet engineering 
is to open topological gaps in the single-particle spectrum. This produces the 
Haldane model\cite{haldane88} using driven honeycomb 
lattices.\cite{eckardt10,jotzu14,goldmanreview14,kitagawa10,zheng14}
Since the possibility to stabilize 
the FCI phases is the ultimate question of this work, one is interested in starting from a 
favorable single-particle bandstructure characterized by relatively flat bands (at least in a 
part of the Brillouin zone) that are separated by large energy gaps. As noted 
previously,\cite{grushin14} this requirement leads to a constraint relating the shaking 
amplitude $\alpha$ to the inverse frequency $\beta$. The detailed discussion of this matter is 
delegated to the Appendix \ref{appa}, with the conclusion that the relative strengths of NNN and NN 
transition amplitudes must be close to the ratio
\begin{equation}
  \label{eq:b2a}
  \frac{\beta \, \rJ_1^2(\alpha)}{\rJ_0 (\alpha)} = \frac{1}{4\sqrt{6}}.
\end{equation} 
Thus, for any given value of the inverse shaking strength $\beta$, the constraint (\ref{eq:b2a}) 
defines the corresponding ``nominal'' value of the driving strength $\alpha_0$ which optimizes 
the single-particle bandstructure. Let us stress, however, that thus defined value $\alpha_0$ 
should be regarded more as \emph{representative} rather than precisely defined optimal value. 
We verified that moderate deviations of the value of $\alpha$ from $\alpha_0$ do not change 
the results qualitatively. Thus, focusing on a single value helps to reduce the dimensionality 
of the parameter phase space.

\subsection{Role of interactions}

A cornerstone of the present contribution is the argument that the real-space micromotion 
couples to the particle interactions through Eq.~(\ref{eq:hf3fin}), which leads to the 
generation of new (and modification of existing\cite{guo15}) interaction terms, and in particular, 
influences the formation and stability of FCI phases. Possible interplay of micromotion and 
interactions was previously discussed\cite{bukov14,verdeny13,dalessiocomment14} basing on 
approximation schemes related to the Magnus expansion.\cite{blanes09} The obtained terms 
are proportional to the inverse driving frequency and typically offer a clear physical 
interpretation (such as the density-assisted tunneling). However, as discussed in 
Ref.~\onlinecite{eckardt15} as well as Ref.~\onlinecite{goldman14}, these terms do not 
influence the spectrum within the order of the approximation.

In the present context of circularly shaken honeycomb lattices, micromotion-induced interaction 
corrections were analyzed and the physical nature of the additional terms was identified in
Ref.~\onlinecite{eckardt15}. Following this work and specializing to moderate driving 
amplitudes, such that Fourier components $|m| \leqslant 1$ are sufficient, we observe that
the overall strength of these terms is set by the prefactor
\begin{equation}
  \frac{U J^2 \rJ_1^2(\alpha)}{(\hbar\omega)^2} =
  U \beta^2 \rJ_1^2(\alpha) = \eta\, U.
\end{equation}
Here we introduced the dimensionless quantity $\eta = \beta^2 \rJ_1^2 (\alpha)$ as a natural 
measure of the relative strength of micromotion-induced interactions with respect to the 
bare onsite repulsion. The five additional contributions for bosons with contact interactions 
read:
\begin{subequations}
\begin{align}
\label{eq:}
  \Ho_F^{(3,\bose)} &= -2z\eta\, U \sum_{i} \no_i \left(\no_i - 1\right) \\
  &+ 4\eta\, U \sum_{\la ij \ra} \no_i \no_j \\
  &+ 2\eta\, U \sum_{\la ij \ra} \aa_i \aa_i \ao_j \ao_j \\
  &- \tfrac{1}{2} \eta\, U \sum_{\la ijk \ra} \aa_i (4\no_j - \no_i -\no_k) \ao_k \\
  &+ \tfrac{1}{2} \eta\, U \sum_{\la ijk \ra} 
    \left(\aa_j \aa_j \ao_i \ao_k + h.c.\right).
\end{align}
\end{subequations}
Here, $z = 3$ is the coordination number (the number of nearest neighbors), and the sums are 
taken, respectively, over the lattice sites $i$, all directed NN links $\la ij \ra$, and all 
directed three-site strings $\la ijk \ra$ with $i$ and $k$ being next-nearest neighbors connected 
via an intermediate site $j$. The physical interpretation of the obtained 
terms is as follows: (a) reduction of the onsite interaction strength, (b) nearest-neighbor
density-density interaction, (c) pair tunneling, (d) density-assisted tunneling between NNN 
sites, and (e) cotunneling of pairs of particles into (from) a given site from (into) two
\emph{distinct} nearest neighbors. While for fermionic particles the corresponding analysis
becomes more involved and is not presented here, we emphasize that the general comparison of 
the relative importance of micromotion-induced interactions and the definition of $\eta$ 
remain unchanged except for the replacement of onsite repulsion energy $U$ with NN repulsion
energy $V$.

In general, the overall scale of the micromotion-induced interactions (i.~e., the value of 
$\eta$) is not an independent parameter but is set by the driving frequency ($\beta$) and 
strength ($\alpha$). Since the preferred value of the driving strength $\alpha_0$ is in 
turn fixed by the flat-band condition, all in all, we have to consider a one-dimensional cut 
through the apparently three-dimensional parameter space $(\alpha, \beta, \eta)$. The 
characteristic range of values is summarized in Table~\ref{tab}.

\begin{table}[ht]
\caption{\label{tab}Numerical values of parameters used in calculation.}
\begin{ruledtabular}
\begin{tabular}{lll}
  $\beta = 0.1$  &  $\alpha_0 = 1.814$  &  $\eta = 0.003$ \\
  $\beta = 0.2$  &  $\alpha_0 = 1.391$  &  $\eta = 0.011$ \\
  $\beta = 0.3$  &  $\alpha_0 = 1.153$  &  $\eta = 0.021$ \\
  $\beta = 0.4$  &  $\alpha_0 = 1.004$  &  $\eta = 0.031$ \\
  $\beta = 0.5$  &  $\alpha_0 = 0.900$  &  $\eta = 0.041$ \\
\end{tabular}
\end{ruledtabular}
\end{table}

\begin{align*}
\end{align*}
As explained previously, the eligible values of the dimensionless inverse shaking frequencies 
$\beta$ are constrained to an interval consistent with the physical setting. One can see, that
the corresponding values of the shaking strength $\alpha_0$ are also restricted to an interval
where: (i) the single-particle bands do not collapse as the zeros of the zeroth-order Bessel
function are avoided, and (ii) the employed truncation of the Fourier series is indeed valid.

Turning to the last column displaying the values of $\eta$, one observes that the relative
contribution of micromotion-induced corrections to the overall strength of interactions is
limited to a few percent. However, these low values should not be taken as a proof that
micromotion-induced interactions may be neglected. In fact, in the context of the 
stabilization of FCI phases, their importance should be measured with respect to manybody
gaps separating the ground-state manifold, and judged by their impact on the FCI stability
diagram. As we will see shortly, micromotion-induced interactions may indeed be significant 
and tend to destabilize fractional Floquet states.

\section{Results}\label{sec:res}
\subsection{Numerical procedure}

To gain specific insights into the impact of micromotion on the formation and stability of FCI 
phases we performed a numerical exact-diagonalization study of periodically driven honeycomb 
lattices. The most stable fractional quantum Hall-like states are expected to form at filling 
fractions $\nu = \tfrac{1}{3}$ for fermions and $\nu = \tfrac{1}{2}$ for bosons. Therefore, 
our numerical modeling mostly focused on finite systems of $N_p = 8$ spinless fermions (resp.\ bosons) 
moving on a lattice containing $N_1 \times N_2 = 6 \times 4$ (resp.\ $N_1 \times N_2 = 4 \times 4$) 
elementary two-site cells. While comparisons with results obtained for larger systems were used 
for consistency checks and validation of the general conclusions, the bulk of calculations was 
performed on moderate eight-particle systems. Keeping the system size relatively small allowed 
for looping over a dense grid in the parameter phase space defined by the aforementioned inverse 
driving frequency $\beta$ and the interaction strength $U$ or $V$, as well as the auxiliary 
phases $\gamma_1$ and $\gamma_2$ introduced through 
twisted boundary conditions\cite{niu84} in the two 
primary lattice directions. Twisting the boundary conditions in the $i$th direction by a phase
factor $\re^{\ri\gamma_i}$ represents the insertion of the dimensionless flux $\gamma_i$, and
leads to the variation of the calculated energy levels, which is commonly referred to as the
spectral flow. The formation of the topological order is signaled (however, not rigorously 
proven) by the formation of the ground-state manifold (GSM) consisting of $\nu^{-1}$ 
quasidegenerate states that rearrange under the spectral flow but remain isolated from the 
other states by a finite manybody gap.\cite{regnault11,bergholtz13}

In view of the translational invariance, restored in the finite system 
by the cyclic boundary conditions, 
our calculations are performed in the reciprocal (or quasimomentum) $\bk$-space. The finite 
geometry of the lattice imposes a discreet grid of $N_1 \times N_2$ permissible values in a 
Brillouin zone (BZ) for both single-particle and total quasimomenta. The diagonalizations can 
be performed separately at each total quasimomentum $\bk$ thus significantly reducing the size of the problem. 
To quote a specific example, in the $8$-fermion system the dimensionality of the Hilbert space 
reduces from $24! / (16! \, 8!) \approx 7.35 \times 10^5$ to just above $3 \times 10^4$ when 
translational invariance is taken into account. The $\bk$-space points belonging to a single 
Brillouin zone are labelled by integer pairs $(k_1, k_2)$ with $k_1 \in [0, N_1-1]$ and 
$k_2 \in [0, N_2-1]$ or, equivalently, by a single index $K = k_1 + N_1 k_2$ with 
$K \in [0, N_1 N_2 - 1]$. The total quasimomentum sectors at which the FCI states will form 
are predicted by a simple counting rule\cite{regnault11} whose validity in the present case 
is indeed confirmed by the actual numerical results. Thus, for eight fermions, the FCI states 
form at $K = \{0, 2, 4\}$, while for eight bosons one finds two states at $K = 0$.

Following previous works that demonstrated the potential existence of FCI phases in various
lattice models\cite{neupert11,sheng11,regnault11,wang11,wu12} with strong particle 
interactions, we use the customary band projection technique (see, e.g., reviews in 
Refs.~\onlinecite{bergholtz13,parameswaran13} for an extensive discussion). Thereby, only 
processes in the relevant single-particle band are included, while the interband scattering 
events as well as processes related to the remaining band are neglected. Note however, that
we \emph{do not} flatten the bandstructure which is another standard tool of the 
trade.\cite{neupert11,bergholtz13} The reason for this choice of techniques is the following:
In contrast to models based on energy bands featuring weak dispersion throughout the Brillouin 
zone, we deal here with single-particle bands that are in general dispersive, however, the upper 
(resp.\ lower) single-particle band develops a relatively flat section at the bottom 
(resp.\ top) where also the Berry curvature is the largest. In this situation, band flattening
would lead to an essential distortion of the model and therefore must be avoided.
When modeling fermionic systems we adopt the approach of Ref.~\onlinecite{grushin14} and 
consider a gas at the density of $4$ particles per $3$ elementary cells. This produces a 
completely filled, and thus inert, lower band plus a partially filled upper band with particles 
predominantly concentrating in the flat section. Therefore, projecting onto the upper band we are 
able to focus on the physics at the effective filling factor $\nu = \tfrac{1}{3}$. 
Proceeding to bosonic systems we envision a gas at the density of one particle per 
$2$ elementary cells, and employ projection onto the lower single-particle band.

\subsection{Manybody topological gap}

As discussed previously, the parameter phase space to be explored is essentially two-dimensional, 
and is spanned by the inverse driving frequency $\beta = J / \hbar\omega$, and the particle 
repulsion strength. It is natural to measure the interaction energies with respect to the basic 
hopping amplitude, thus introducing the corresponding dimensionless quantities
\begin{equation}
\label{eq:udim}
  u = \frac{U}{J}, \quad\text{and}\quad v = \frac{V}{J},
\end{equation}
for bosonic and fermionic systems, respectively.

\begin{figure}[ht]
\begin{center}
\includegraphics[width=84mm]{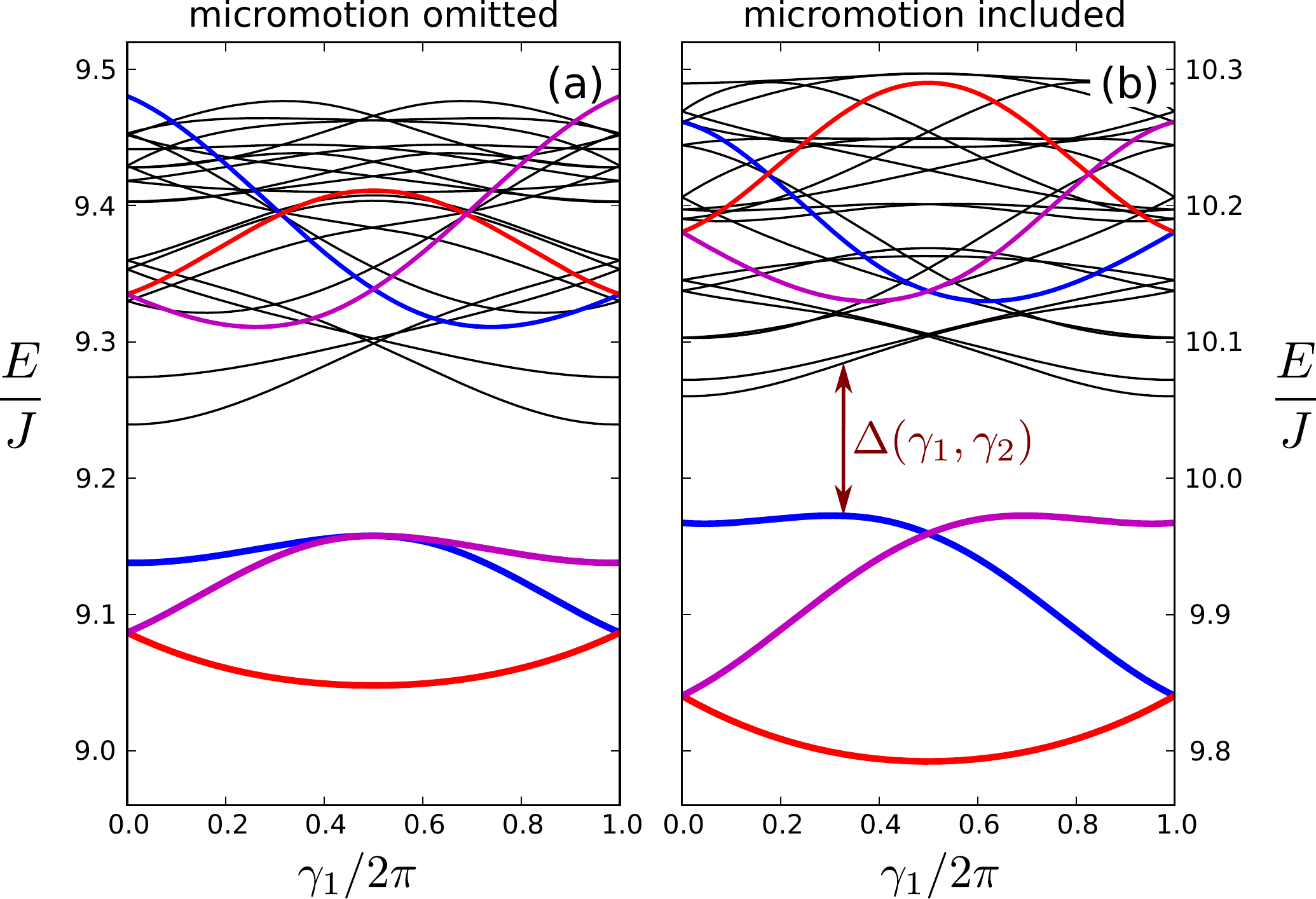}
\caption{\label{fig:spectralflow}
Typical energy spectra of the driven honeycomb lattice with micromotion-induced interactions 
ignored [panel (a)], and taken into account [panel (b)] plotted versus the auxiliary
flux $\gamma_1$ inserted along the first (commensurate) lattice direction. Here, the energies 
of manybody states in the eight-fermion system numerically calculated at $\beta = 0.3$, $v = 5$, 
and $\gamma_2 = 0$ are shown. Colored (resp.\ black) lines denote the states in the ground-state
manifold 
(resp.\ other) total quasimomentum sectors, and two lowest-energy states of each quasimomentum 
sector are included. The dark red arrow illustrates the definition of the \emph{local} manybody 
gap referred to in the text.}
\end{center}
\end{figure}

Figure~\ref{fig:spectralflow} shows the calculated energy spectra of eight-fermion manybody 
states in the form of spectral flows along the first (commensurate) direction. The three 
states belonging to the expected FCI total quasimomentum sectors $K = \{0, 2, 4\}$ are plotted in color 
while the remaining states are plotted in black. The insertion of the artificial flux 
$\gamma_1$ defining the twisted boundary conditions makes the three states in the GSM 
interchange and reconnect to their partners at the opposite boundary of the manybody Brillouin 
zone (MBZ). The plot is obtained by setting $\beta = 0.3$, $v = 5$, and $\gamma_2 = 0$. In the 
left panel, the micromotion strength $\eta$ is artificially set to zero (as would happen in a 
calculation omitting the presence of real-space micromotion). The nearly degenerate states in the GSM 
stay protected from the excited states by a manybody gap, and their total Chern number obtained 
from sampling over the whole MBZ $(\gamma_1, \gamma_2) \in [0,2\pi) \times [0,2\pi)$ sums up 
to unity with numerical precision. The right panel shows the realistic situation where coupling 
between micromotion and interactions is duly taken into account up to the third order. While the 
topological nature of the lowest-energy manifold and the manybody gap persist, its width is reduced 
and the energy spread of the states in the GSM is now considerably larger. These observations 
serve as an early hint that the interplay of micromotion and interactions is significant, and
may indeed have a detrimental role on the stability of the Floquet fractional Chern insulating
phases.

Referring to Fig.~\ref{fig:spectralflow}, we also take the opportunity to give a precise
definition to the notion of the manybody gap to be used in presentation of further results. 
Since calculations are performed looping over the manybody Brillouin zone 
we first define the \emph{local} (that is, corresponding to fixed values of $\gamma_{1,2}$)
dimensionless manybody gap 
\begin{equation}
  \Delta (\gamma_1, \gamma_2) = \frac{1}{J} \left[\operatorname*{min}_{\text{not in GSM}} 
  E(\gamma_1, \gamma_2)
  - \operatorname*{max}_{\text{GSM}} 
  E(\gamma_1, \gamma_2)\right],
\end{equation}
as the energy difference between the highest state belonging to GSM and the lowest-lying state 
outside the GSM measured in units of $J$. Thus, $\Delta (\gamma_1, \gamma_2)$ is positive when 
an isolated ground state manifold is formed and negative when the states in the ground-state 
manifold mix with the remaining states. The overall manybody gap is then obtained by minimizing 
the local manybody gap over the entire MBZ, thus
\begin{equation}
  \Delta = \operatorname*{min}_{\text{MBZ}} \Delta (\gamma_1, \gamma_2).
\end{equation}

\begin{figure}[ht]
\begin{center}
\includegraphics[width=84mm]{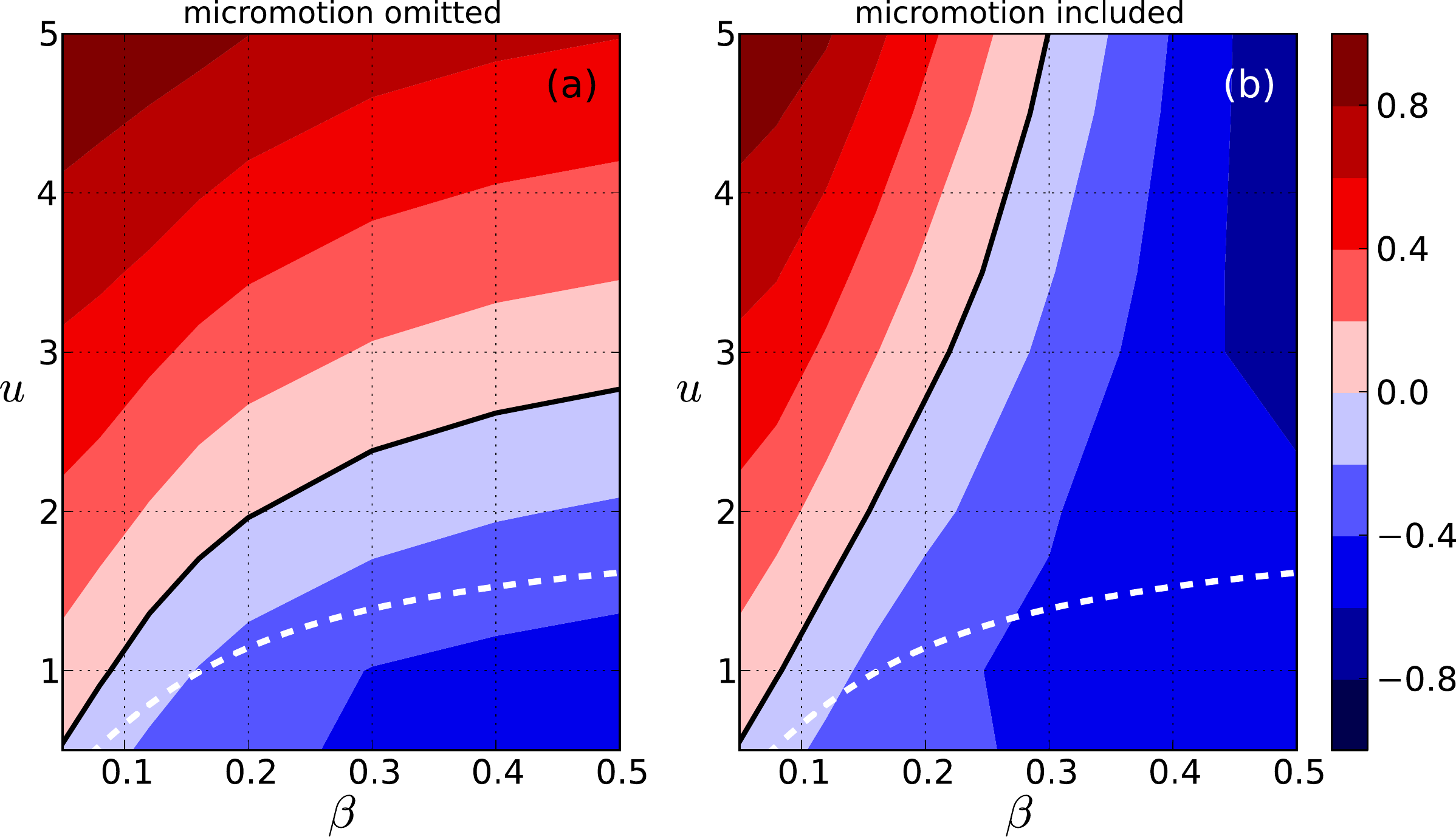}
\caption{\label{fig:boson}
The phase diagram showing the manybody gap $\Delta$ as a function of the inverse shaking 
frequency $\beta$ and the dimensionless interaction strength $u$ for eight-boson system. 
Micromotion-induced interactions are omitted in panel (a) and taken into account to the third
order in panel (b). The full black lines delimit the regions of positive manybody gaps
(shown in red shades) and differ considerably between the panels. The white dashed line
indicates the parameter regime where the interaction strength $U$ is equal to the 
single-particle bandgap. Well below this line mixing between single-particle bands becomes small.}
\end{center}
\end{figure}

\begin{figure}[ht]
\begin{center}
\includegraphics[width=84mm]{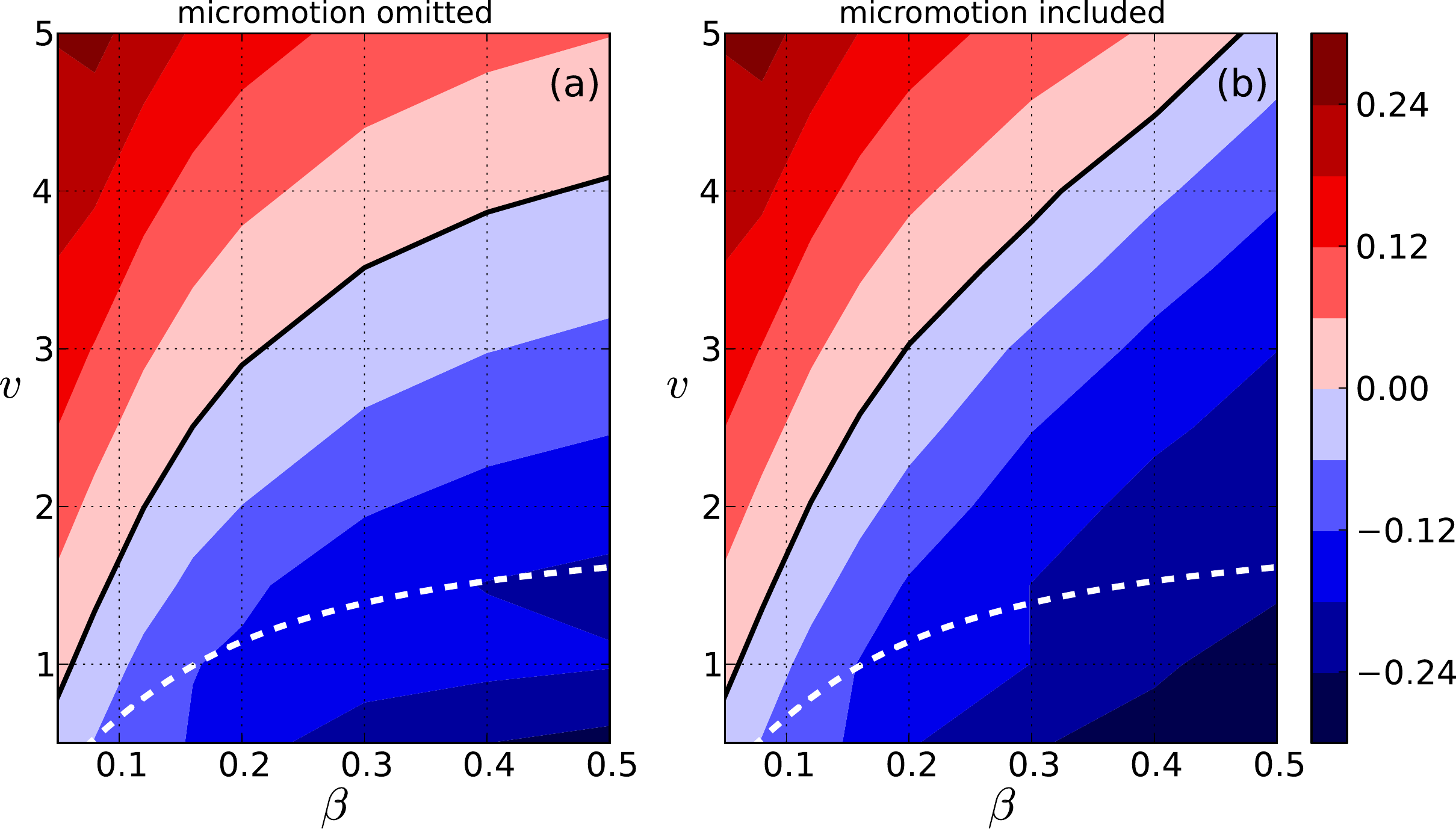}
\caption{\label{fig:fermion}
The same as Fig.~\ref{fig:boson} but for eight-fermion system with dimensionless NN repulsion
energy $v$. Note that the scale of the colorbar, and correspondingly the typical manybody gaps 
are significantly smaller than for bosons.}
\end{center}
\end{figure}

The complete phase diagrams in the $\beta$-$u$ (resp.\ $\beta$-$v$) plane for bosons (resp.\ fermions) 
occupying the lower (resp.\ upper) single-particle band at $1/3$ (resp.\ $1/2$) filling are shown in 
Figs.~\ref{fig:boson} and \ref{fig:fermion}. In both instances we map out the behavior of the 
manybody gap as a function of the governing parameters, $\beta$ and $u$ (resp.\ $v$). The red areas 
correspond to the presence of a positive manybody gap implying that the states comprising 
the GSM are isolated from the remaining ones \emph{everywhere} in the MBZ and thus the two 
manifolds do not mix. On the contrary, blue areas correspond to a \emph{negative} manybody 
gap. In this situation, at least one state from the GSM attains higher energy value than the 
lowest state not belonging to the GSM, and the two manifolds overlap somewhere in the MBZ. The 
full black line separates phase space areas characterized by manybody gaps of different signs 
and thus serves as a first criterion for the possibility of formation of the Floquet fractional
Chern insulator. Of course, this criterion takes into account only the aspects of the energy 
spectra and must be further supported by, e.~g., information obtained from excitation spectra
discussed in the following subsection. Nevertheless, already these phase diagrams indicate that 
inclusion of micromotion-induced interactions typically pushes the phase boundary 
upwards in the phase diagram. That is, stronger interactions will be needed to stabilize FCI 
phases at equal other conditions.

Bosonic lattice systems seem to be much more promising candidates for the stabilization of FCI 
phases than fermionic ones. Indeed, they find a realization in quantum gas experiments where
strong and tunable onsite interactions are possible, and Fig.~\ref{fig:boson} reveals that in
bosonic systems 
large manybody gaps on the order of the tunneling amplitude $J$ can be obtained. In contrast, 
for fermionic systems nearest-neighbor repulsions are harder to tune into strongly-interacting
limit (in graphene the strength of NN repulsion is estimated\cite{wehling11} to be around $2J$), 
and the overall scale of attainable manybody gaps is also smaller, 
as demonstrated by Fig.~\ref{fig:fermion}.

Viewing FCI states as direct analogues of Laughlin's fractional quantum Hall states, one would 
ideally prefer working in the regime where the fractional states are induced by strong 
interactions mixing single-particle states within a single topological (Chern or Landau) band 
and not among several bands. (Recent results suggest, however, that weak interband mixing 
is not detrimental for the formation of fermionic $\nu = \tfrac{1}{3}$ FCI state\cite{kourtis13,gmzp15}). 
This corresponds to the situation where the single-particle bandgap exceeds the characteristic 
interaction strength. In Figs.~\ref{fig:boson} and \ref{fig:fermion}, the white dashed lines
indicate the regime where the corresponding interaction parameter, $U$ or $V$, is equal to the 
single-particle bandgap; well below these lines mixing between single-particle bands is small.
Obviously, the approximation based on the band projection is not justified in the considered 
regime. Moreover, the interaction strengths required to stabilize FCI phases in the current model 
also generally exceed the above-stated limit. These observations suggest that it must be
quite challenging to explore FCI phases within the considered model. A window of
opportunity might persist at the higher end of the considered range of driving frequencies
(equivalently, at the lower end of the range of $\beta$'s). Although the approximation focusing
on just one single-particle band is not fully justified here, the effects of band mixing were 
reported not to be critical.\cite{kourtis13,gmzp15} In this limit, notably weaker interaction 
strengths are required to stabilize FCI phases [see Figs.~\ref{fig:boson} and \ref{fig:fermion}], 
and moreover, problems caused by heating due to the resonant excitation of collective excitations 
become smaller.

\subsection{Quasihole spectra}

\begin{figure}[ht]
\begin{center}
\includegraphics[width=84mm]{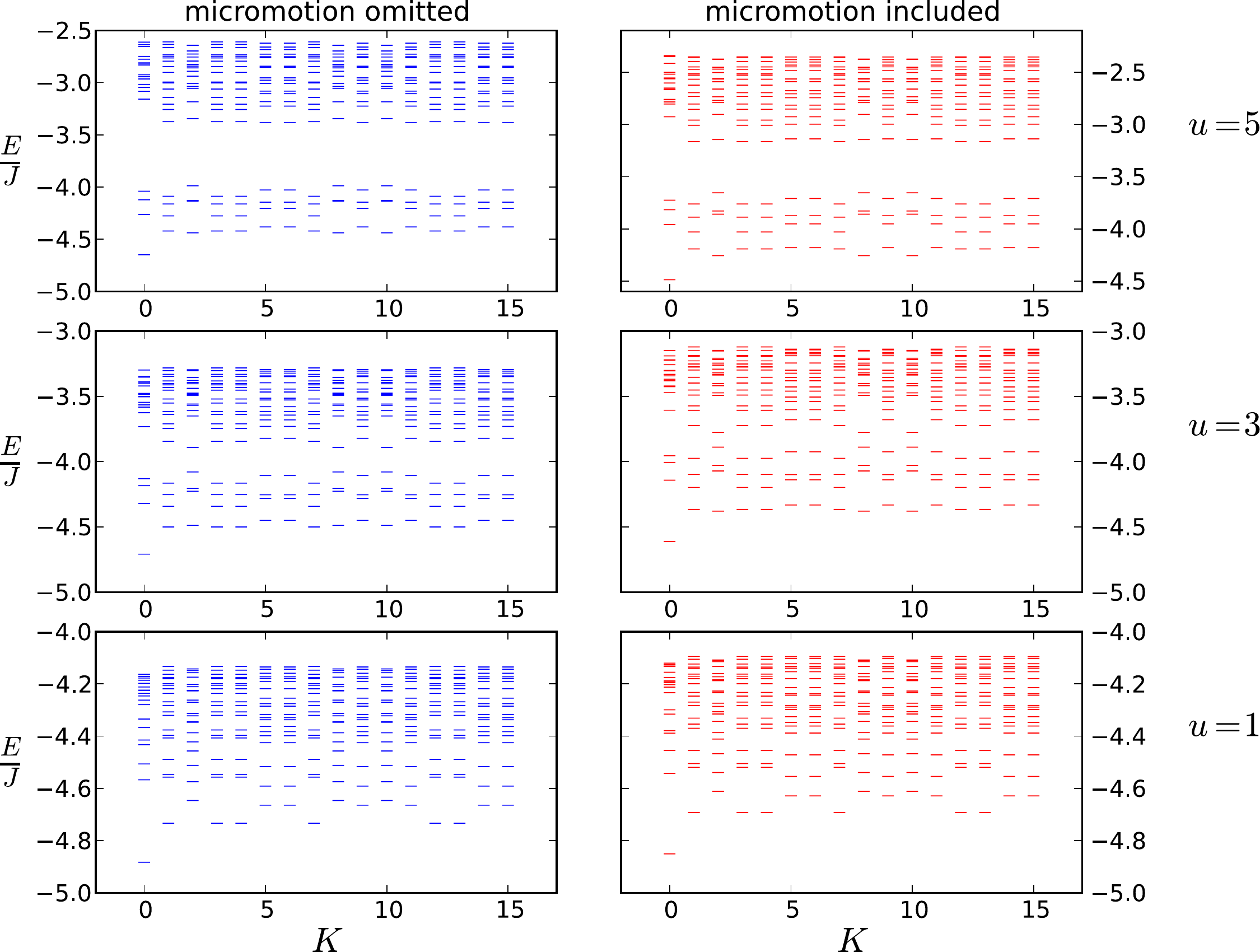}
\caption{\label{fig:qhboson}
Quasihole excitation spectra for eight-boson system calculated at the inverse dimensionless driving 
strength $\beta = 0.1$. Formation of an isolated manifold of $4\times 16$ states is visible for 
strong interactions with $u = 5$, and gradually closes when interaction become weaker.}
\end{center}
\end{figure}

Finally, let us also take a brief look at the quasihole excitation spectra, corresponding to the
removal of a particle from the ground state.\cite{regnault11} Here, we are driven by twofold 
motivation. On the one hand, these spectra serve as an additional identification check confirming 
that the obtained phases display behavior characteristic of fractional Chern insulator. Thus, 
one expects to observe the formation of an isolated low-energy manifold of quasihole states 
whose number is predicted by counting rules based on the generalized Pauli principle.\cite{regnault11} 
On the other hand, one is interested in the the evolution of the size of the gap separating the 
low-energy manifold from the remaining states as a function of the computational parameters.
In this way, we also obtain complementary information about the interaction strengths that are 
sufficient to induce the FCI phases.

Qualitatively similar results are obtained for both fermionic and bosonic systems. Focusing on 
the latter, we perform exact diagonalizations for systems of $7$ bosonic particles moving on a 
lattice consisting of $4 \times 4$ elementary cells. 
This is one particle less than would correspond to the exact $1/2$ filling
of the band, and can be interpreted as an introduction of a hole into the previously studied system. 
In this specific case, the $(1,2)$-admissible counting rule\cite{regnault11} predicts the 
formation of an isolated manifold with $4$ states per each total quasimomentum sector, implying
the total of $64$ states. The actual numerical results obtained for $\beta = 0.1$ and varying values
of the interaction strength are shown in Fig.~\ref{fig:qhboson}. In all panels, the abscissa
axes enumerate the total quasimomentum sectors indexed by the integer
values $K$ running from zero to $15 \equiv 4\times 4 -1$, and the ordinate axes display the 
energies of the lowest manybody states obtained from the exact-diagonalization calculations. 
The left (resp.\ right) column of three plots corresponds to omitted (resp.\ included to the 
third order) micromotion-induced corrections to the effective Hamiltonian. The strength of 
particle interactions $u = U / J$ is decreasing from the top to the bottom. At the strongest 
considered interactions, $u = 5$ (top row) one sees a clear gap separating the lower group of 
exactly four states per quasimomentum sector from the rest of the spectrum. For weaker interactions 
($u = 3$ in the middle row) the gap is barely discernible in the absence of micromotion-induced
interactions and is obliterated when micromotion is taken into account. Finally, the in bottom
row corresponding to weak interactions with $u = 1$ there is no visible gap indicating
the absence of FCI phase. In general, the information provided by the quasihole spectra is
broadly compatible with that obtained from the ordinary manybody spectra, however,
the constraints placed on interactions being ``sufficiently strong'' are even more stringent,
thus further contributing to the pessimistic outlook on the feasibility of fractional Chern 
insulating states in the studied system.

\section{Conclusions}\label{sec:con}

To summarize, we address the issue of the stabilization of Floquet fractional Chern insulator 
states for strongly interacting particles (bosons or fermions) on a time-periodically driven 
honeycomb lattice. In this system, the necessary topological single particle bands are formed 
due to the next-nearest neighbor transitions, which is a second-order effect corresponding to 
two consecutive tunneling events during a single driving period. This requires sufficiently low 
driving frequencies and necessitates the consideration of further expansion terms beyond the
averaging of the Hamiltonian used in high-frequency schemes. The third-order terms describe the
coupling of real-space micromotion and interactions. In strongly interacting systems, the importance 
of these terms is, in general, comparable to that of the second-order contributions. The prefactor
features not only the expansion parameter $(\hbar\omega)^{-1}$ but also the onsite repulsion 
energy $U$ for bosonic systems or, alternatively, nearest-site repulsion energy $V$ for fermionic 
systems. Within simulations of small systems, the coupling of micromotion and interactions turns 
out to be both significant and detrimental to the formation of quantum-Hall like states. Thus, 
the realization of Floquet fractional Chern insulator states seems rather challenging.

\acknowledgments

The authors thank T.~Neupert and A.~G.~Grushin for valuable discussions and comments on a
preliminary version of the manuscript. This work was supported by the European Social Fund 
under the Global Grant measure. B.~A.\ was supported by the the ARO Atomtronics MURI and the 
NSF under Grant No.~PHY-100497 and the Physics Frontiers Center Grant PHY-0822671.

\appendix
\section{Haldane model}\label{appa}

The Haldane model\cite{haldane88,shao08,stanescu10,alba11,hauke12,goldman13,anisimovas14,jotzu14,baur14} 
sets a paradigmatic example of the topological 
bandstructure supported by a simple two-band setup. This model features a honeycomb lattice 
(see Fig.~\ref{fig:latt}) with nearest-neighbor hopping described by a real amplitude $-J_1$ 
plus next-nearest neighbor hopping described by a \emph{complex} amplitude 
$-J_2\,\re^{\ri\varphi}$ (resp.\ $-J_2\,\re^{-\ri\varphi}$) in the counter-clockwise 
(resp.\ clockwise) direction. In the quasimomentum representation, the Hamiltonian matrix reads
\begin{equation}
  {H} (\bk) = \begin{bmatrix}
  -J_2 \tilde{f} (\bk, \varphi) & - J_1 g^* (\bk) \\
  - J_1 g (\bk) &
  -J_2 \tilde{f}(\bk, -\varphi) 
  \end{bmatrix},
\end{equation}
with
\begin{align}
  g (\bk) &= \sum_{j=1}^3 \re^{-\ri\bk\cdot(\bm\delta_j-\bm\delta_1)}
    = 1 + \re^{\ri (k_1 + k_2)} + \re^{\ri k_2}, \label{app:a2}\\
  \tilde{f} (\bk, \varphi) 
    &= 2 \, \sum_{j=1}^3 \cos \left(\bk\cdot \ba_j - \varphi \right) \nonumber\\
    &= 2 \, \big[\cos(k_1 - \varphi) + \cos(k_2 - \varphi) \nonumber\\
    &+ \cos(k_1 + k_2 + \varphi)\big]. \label{app:a3}
\end{align}
Here we define $\bk = (k_1/2\pi)\,\bm{b}_1 + (k_2/2\pi)\, \bm{b}_2$, with $\bm{b}_{1,2}$ denoting
the reciprocal lattice vectors, and $(k_1, k_2) \in [0, 2\pi) \times [0, 2\pi)$ covering the 
rhombic Brillouin zone. 
In the case considered in the main text, the NNN hopping amplitude is purely imaginary, 
corresponding to $\varphi \to \pi/2$, and therefore
\begin{equation}
  {H} (\bk) = \begin{bmatrix}
  - J_2 f (\bk) & - J_1 g^* (\bk) \\
  - J_1 g (\bk) &
  J_2 f(\bk) 
  \end{bmatrix},
\end{equation}
with the unchanged definition for $g(\bk)$, and 
\begin{equation}
\begin{split}
  \label{app:a5}
  f (\bk) = 2 \, \big[\sin k_1 + \sin k_2 - \sin (k_1+k_2)\big].
\end{split}
\end{equation}
While the resulting bands are not globally flat, they feature flat sections that cover BZ 
regions characterized by large Berry curvatures. This fact forms the basis for the 
anticipation\cite{grushin14} that the Haldane model might support FCI phases for 
strongly interacting particles.

\begin{figure}[ht]
\begin{center}
\includegraphics[width=84mm]{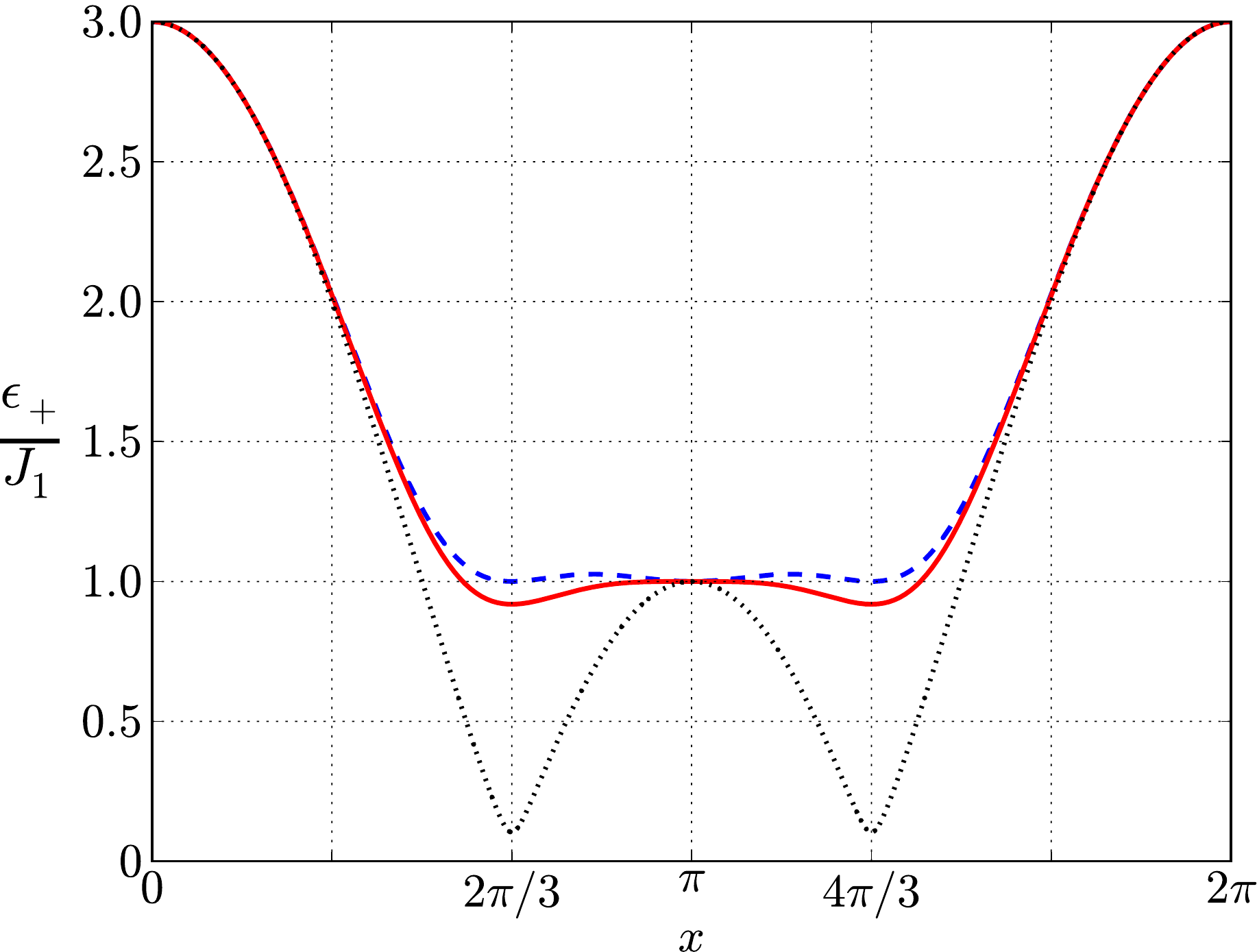}
\caption{\label{fig:haldaned}Dispersion of the upper band of the Haldane model drawn along the 
diagonal of the rhombic BZ.}
\end{center}
\end{figure}

The formation of the flat segments in the energy bands is best visualized focusing on a 
one-dimensional cut through the rhombic BZ along its diagonal (obtained by setting 
$k_1 = k_2$). This line connects 
the zone center at $(0,0)$ to the equivalent point at $(2\pi, 2\pi)$, and passes through both 
inequivalent Dirac points at $(2\pi/3, 2\pi/3)$ and $(4\pi/3, 4\pi/3)$. Using the single 
coordinate, $x \equiv k_1 = k_2$, we simplify (\ref{app:a2}) and (\ref{app:a5}) to read
\begin{align}
  g (x) &= 1 + \re^{i x} + \re^{i 2x},\\
  f (x) &= 4 \sin x - 2 \sin 2x,
\end{align}
and obtain the scaled band energies
\begin{equation}
  \frac{\epsilon_{\pm}}{J_1} = \pm \left[\left(\frac{J_2}{J_1}\right)^2 f^2(x) + |g(x)|^2\right]^{1/2}.
\end{equation}
The two bands are mirror-symmetric with respect to the line $\epsilon = 0$, and
the dispersion of the upper band is shown in Fig.~\ref{fig:haldaned} for a few selected values 
of the governing parameter $F = J_2 / J_1$. Weak NNN hopping amplitudes open small topological 
gaps at the Dirac points, as illustrated by the dotted black line corresponding to $F = 0.02$. 
Stronger couplings produce relatively dispersionless portions of the bands. 
In Fig.~\ref{fig:haldaned}, this result is illustrated by the full red and dashed blue lines 
drawn, respectively for $F_Q = 1 / \sqrt{32}$ and $F_D = 1/\sqrt{27}$. The two reference
values are obtained from different flatness criteria, and are very similar. The first value,
$F_Q$, is defined by the requirement that the Taylor expansion of the energy $\epsilon_{+}/J$ 
around the midpoint $x = \pi$ starts with the quartic (rather than the second order) term. The 
latter value, $F_D$, is the smallest at which the bandgap at the Dirac points becomes equal 
to the gap at the midpoint of the BZ diagonal. These two alternative criteria lead only to 
miniscule variations in the resulting manybody bandstructure, and we choose to adopt the 
value $F_Q$ as the reference.

\section{Operators in quasimomentum representation}\label{appb}

In the main text, we discuss the constituent parts of the effective Hamiltonian using their
real-space forms which typically lend themselves to physically transparent interpretation as
combinations of hopping and interaction events. The purpose of the present Appendix is to give
the corresponding expressions in the reciprocal space, used in the actual numerical work.

Denoting the number of elementary cells in the lattice $N_s$, we define the reciprocal-space 
creation operators $\aa_{\bk \rA}$ and $\aa_{\bk \rB}$ in terms of the usual Fourier transforms
\begin{align}
  \aa_{\bk \rA} &= \frac{1}{\sqrt{N_s}} \sum_{i \in \rA}
    \aa_{i} \, \re^{\ri \bk\cdot\br_i}, \\
  \aa_{\bk \rB} &= \frac{1}{\sqrt{N_s}} \sum_{i \in \rB}
    \aa_{i} \, \re^{\ri \bk\cdot(\br_i-\bm\delta_1)},
\end{align}
with the conjugate version of this equation applicable to annihilation operators. Note that in 
the quasimomentum representation we explicitly specify the sublattice index $\rA$ or $\rB$ while 
in the real space this was avoided by using the shorthand notation $i$ versus $i+\mu$.

The first-order contribution to the effective Hamiltonian is now written
\begin{equation}
  \Ho_F^{(1)} = \Ho_{\rm int} - J \rJ _0 (\alpha) \sum_{\bk}
  \left[\aa_{\bk \rB} \ao_{\bk \rA} \, g(\bk) + h.c. \right],
\end{equation}
with $g(\bk) = \sum_{\mu} \re^{-i\bk\cdot(\bm\delta_\mu - \bm\delta_1)}$. 
The interaction Hamiltonian reads for bosons
\begin{subequations}
\begin{align}
  \Ho_{\rm int}^{(\bose)} &= \frac{1}{2} \sum_{\kkkk} W_{\bose}\left(\kkkk \right)
  \Big[ \aa_{\bk_1 \rA} \aa_{\bk_2 \rA} \ao_{\bk_3 \rA} \ao_{\bk_4 \rA} \nonumber\\
  &+ \aa_{\bk_1 \rB} \aa_{\bk_2 \rB} \ao_{\bk_3 \rB} \ao_{\bk_4 \rB} \Big], \\ 
  W_{\bose} \left(\kkkk \right) 
    &= \frac{U}{N_s} \, \delta'_{\bk_1+\bk_2, \bk_3+\bk_4},
\end{align}
\end{subequations}
and for fermions
\begin{subequations}
\begin{align}
  \Ho_{\rm int}^{(\fermi)} &= \sum_{\kkkk}
    \aa_{\bk_1 \rB} \aa_{\bk_2 \rA} \ao_{\bk_3 \rA} \ao_{\bk_4 \rB}
    \cdot W_{\fermi}\left(\kkkk \right), \\ 
  W_{\fermi} \left(\kkkk \right) 
    &= \frac{V}{N_s} \, g(\bk_1 - \bk_4) \, \delta'_{\bk_1+\bk_2, \bk_3+\bk_4},
\end{align}
\end{subequations}
with the periodic Kronecker delta that allows the quasimomenta in its arguments to differ by an 
elementary translation on the reciprocal lattice. 
The symbol $\{\bk\} \equiv (\bk_1, \bk_2, \bk_3, \bk_4)$ stands for the set of four quasimomenta 
involved in a scattering event.

Proceeding to the second order we write
\begin{equation}
  \Ho_1 = -J \rJ_1 (\alpha) \sum_{\bk} \left[
  \aa_{\bk \rB} \ao_{\bk \rA} \hmn{\rB\rA}{\bk} + \aa_{\bk \rA} \ao_{\bk \rB} \hmn{\rA\rB}{\bk},
  \right]
\end{equation}
with
\begin{subequations}
\begin{align}
  \hmn{\rB\rA}{\bk} &= \sum_{\mu=1}^3 
    \re^{-\ri \bk\cdot (\bm\delta_{\mu} - \bm\delta_1)} \re^{-\ri\varphi_{\mu}},\\
  \hmn{\rA\rB}{\bk} &= -\sum_{\mu=1}^3 
    \re^{\ri \bk\cdot (\bm\delta_{\mu} - \bm\delta_1)} \re^{-\ri\varphi_{\mu}},
\end{align}
\end{subequations}
and evaluating the commutator in Eq.~(\ref{eq:hamfb}) obtain
\begin{equation}
\begin{split}
  \Ho_F^{(2)} &= \frac{1}{\hbar\omega} \left[\Ho_1, \Ho_1^{\dag}\right] 
  = \frac{\sqrt{3} J^2 \rJ_1^2 (\alpha)}{\hbar\omega} \\
  &\times \sum_{\bk} f(\bk) \left[\aa_{\bk\rA} \ao_{\bk\rA} - \aa_{\bk\rB} \ao_{\bk\rB} \right],
\end{split}
\end{equation}
with $f (\bk) = 2 \sum_j \sin \bk\cdot\ba_j$. These terms are diagonal and describe NNN hopping
on separate sublattices.

To describe the coupling of micromotion and interactions, we evaluate the nested commutators in 
Eq.~(\ref{eq:hf3fin}) and obtain, depending on the statistics,
\begin{equation}
  \Ho_F^{(3,\bose|\fermi)} = - \frac{J^2 \rJ_1^2 (\alpha)}{2 \left(\hbar\omega\right)^2}
  \sum_{\kkkk} W_{\bose|\fermi} (\kkkk) \, \mathcal{S}_{\bose|\fermi},
\end{equation}
with
\begin{widetext}
\begin{equation}
\begin{split}
\mathcal{S}_{\bose} (\kkkk) =
&+\hmn{\rA\rB}{\bk_3}\, \hmnconj{\rA\rB}{\bk_3}\, \cfour{\bk_1\rA}{\bk_2\rA}{\bk_3\rA}{\bk_4\rA} 
 -\hmn{\rA\rB}{\bk_3}\, \hmnconj{\rA\rB}{\bk_1}\, \cfour{\bk_1\rB}{\bk_2\rA}{\bk_3\rB}{\bk_4\rA} \\
&-\hmn{\rA\rB}{\bk_3}\, \hmnconj{\rA\rB}{\bk_2}\, \cfour{\bk_1\rA}{\bk_2\rB}{\bk_3\rB}{\bk_4\rA} 
 +\hmn{\rA\rB}{\bk_3}\, \hmnconj{\rB\rA}{\bk_4}\, \cfour{\bk_1\rA}{\bk_2\rA}{\bk_3\rB}{\bk_4\rB} \\
&+\hmn{\rA\rB}{\bk_4}\, \hmnconj{\rA\rB}{\bk_4}\, \cfour{\bk_1\rA}{\bk_2\rA}{\bk_3\rA}{\bk_4\rA} 
 -\hmn{\rA\rB}{\bk_4}\, \hmnconj{\rA\rB}{\bk_1}\, \cfour{\bk_1\rB}{\bk_2\rA}{\bk_3\rA}{\bk_4\rB} \\
&-\hmn{\rA\rB}{\bk_4}\, \hmnconj{\rA\rB}{\bk_2}\, \cfour{\bk_1\rA}{\bk_2\rB}{\bk_3\rA}{\bk_4\rB} 
 +\hmn{\rA\rB}{\bk_4}\, \hmnconj{\rB\rA}{\bk_3}\, \cfour{\bk_1\rA}{\bk_2\rA}{\bk_3\rB}{\bk_4\rB} \\
&+\hmn{\rB\rA}{\bk_1}\, \hmnconj{\rA\rB}{\bk_2}\, \cfour{\bk_1\rB}{\bk_2\rB}{\bk_3\rA}{\bk_4\rA}
 -\hmn{\rB\rA}{\bk_1}\, \hmnconj{\rB\rA}{\bk_3}\, \cfour{\bk_1\rB}{\bk_2\rA}{\bk_3\rB}{\bk_4\rA} \\
&-\hmn{\rB\rA}{\bk_1}\, \hmnconj{\rB\rA}{\bk_4}\, \cfour{\bk_1\rB}{\bk_2\rA}{\bk_3\rA}{\bk_4\rB}
+ \hmn{\rB\rA}{\bk_1}\, \hmnconj{\rB\rA}{\bk_1}\, \cfour{\bk_1\rA}{\bk_2\rA}{\bk_3\rA}{\bk_4\rA} \\
&+\hmn{\rB\rA}{\bk_2}\, \hmnconj{\rA\rB}{\bk_1}\, \cfour{\bk_1\rB}{\bk_2\rB}{\bk_3\rA}{\bk_4\rA} 
 -\hmn{\rB\rA}{\bk_2}\, \hmnconj{\rB\rA}{\bk_3}\, \cfour{\bk_1\rA}{\bk_2\rB}{\bk_3\rB}{\bk_4\rA} \\
&-\hmn{\rB\rA}{\bk_2}\, \hmnconj{\rB\rA}{\bk_4}\, \cfour{\bk_1\rA}{\bk_2\rB}{\bk_3\rA}{\bk_4\rB}
 +\hmn{\rB\rA}{\bk_2}\, \hmnconj{\rB\rA}{\bk_2}\, \cfour{\bk_1\rA}{\bk_2\rA}{\bk_3\rA}{\bk_4\rA}, \\
&+ h.c. + \Big\{ \rA \leftrightarrow \rB \Big\},
\end{split}
\end{equation}

\begin{equation} 
\begin{split}  
  \mathcal{S}_{\fermi} (\kkkk) = 
  &- h_{\rB\rA} (\bk_4) \, h_{\rA\rB}^* (\bk_2) \, \aa_{\bk_1\rB} \aa_{\bk_2\rB} \ao_{\bk_3\rA} \ao_{\bk_4\rA}
   - h_{\rB\rA} (\bk_2) \, h_{\rA\rB}^* (\bk_4) \, \aa_{\bk_1\rB} \aa_{\bk_2\rB} \ao_{\bk_3\rA} \ao_{\bk_4\rA} \\
  &- h_{\rB\rA} (\bk_4) \, h_{\rB\rA}^* (\bk_1) \, \aa_{\bk_1\rA} \aa_{\bk_2\rA} \ao_{\bk_3\rA} \ao_{\bk_4\rA}
   + h_{\rB\rA} (\bk_4) \, h_{\rB\rA}^* (\bk_3) \, \aa_{\bk_1\rB} \aa_{\bk_2\rA} \ao_{\bk_3\rB} \ao_{\bk_4\rA} \\
  &+ h_{\rB\rA} (\bk_4) \, h_{\rB\rA}^* (\bk_4) \, \aa_{\bk_1\rB} \aa_{\bk_2\rA} \ao_{\bk_3\rA} \ao_{\bk_4\rB}
   + h_{\rB\rA} (\bk_2) \, h_{\rB\rA}^* (\bk_1) \, \aa_{\bk_1\rA} \aa_{\bk_2\rB} \ao_{\bk_3\rA} \ao_{\bk_4\rB} \\
  &+ h_{\rB\rA} (\bk_2) \, h_{\rB\rA}^* (\bk_2) \, \aa_{\bk_1\rB} \aa_{\bk_2\rA} \ao_{\bk_3\rA} \ao_{\bk_4\rB}
   - h_{\rB\rA} (\bk_2) \, h_{\rB\rA}^* (\bk_3) \, \aa_{\bk_1\rB} \aa_{\bk_2\rB} \ao_{\bk_3\rB} \ao_{\bk_4\rB} \\
  &- h_{\rA\rB} (\bk_3) \, h_{\rA\rB}^* (\bk_2) \, \aa_{\bk_1\rB} \aa_{\bk_2\rB} \ao_{\bk_3\rB} \ao_{\bk_4\rB}
   + h_{\rA\rB} (\bk_3) \, h_{\rA\rB}^* (\bk_3) \, \aa_{\bk_1\rB} \aa_{\bk_2\rA} \ao_{\bk_3\rA} \ao_{\bk_4\rB} \\
  &+ h_{\rA\rB} (\bk_3) \, h_{\rA\rB}^* (\bk_4) \, \aa_{\bk_1\rB} \aa_{\bk_2\rA} \ao_{\bk_3\rB} \ao_{\bk_4\rA}
   + h_{\rA\rB} (\bk_1) \, h_{\rA\rB}^* (\bk_1) \, \aa_{\bk_1\rB} \aa_{\bk_2\rA} \ao_{\bk_3\rA} \ao_{\bk_4\rB} \\
  &+ h_{\rA\rB} (\bk_1) \, h_{\rA\rB}^* (\bk_2) \, \aa_{\bk_1\rA} \aa_{\bk_2\rB} \ao_{\bk_3\rA} \ao_{\bk_4\rB}
   - h_{\rA\rB} (\bk_1) \, h_{\rA\rB}^* (\bk_4) \, \aa_{\bk_1\rA} \aa_{\bk_2\rA} \ao_{\bk_3\rA} \ao_{\bk_4\rA} \\
  &- h_{\rA\rB} (\bk_3) \, h_{\rB\rA}^* (\bk_1) \, \aa_{\bk_1\rA} \aa_{\bk_2\rA} \ao_{\bk_3\rB} \ao_{\bk_4\rB}
   - h_{\rA\rB} (\bk_1) \, h_{\rB\rA}^* (\bk_3) \, \aa_{\bk_1\rA} \aa_{\bk_2\rA} \ao_{\bk_3\rB} \ao_{\bk_4\rB} \\
  &+ h.c..
\end{split}
\end{equation}
\end{widetext}

\bibliography{mint}

\end{document}